\documentclass{aa}
\usepackage{graphicx}
\usepackage{natbib}
\usepackage{txfonts}
\usepackage{color}
\bibpunct{(}{)}{;}{a}{}{,} 
\begin{document}
   \title{Signature of mass supply to quiet coronal loops}
   \author{H. Tian\inst{1,2}
           \and
           C.-Y. Tu\inst{1,2}
          \and
           E. Marsch\inst{1}
           \and
           J.-S. He\inst{2}
           \and
           G.-Q. Zhou\inst{2}
          }
   \offprints{H. Tian}
   \institute{Max-Planck-Institut f\"ur Sonnensystemforschung, Katlenburg-Lindau, Germany\\
              \email{tianhui924@163.com, marsch@mps.mpg.de}
             \and
             Department of Geophysics, Peking University, Beijing, China\\
             \email{chuanyitu@pku.edu.cn}
             }
   \date{}

\abstract {} {The physical implication of large blue shift of
Ne~{\sc{viii}} in the quiet Sun region is investigated in this
paper.} {We compare the significant Ne~{\sc{viii}} blue shifts,
which are visible as large blue patches on the Doppler-shift map of
a middle-latitude quiet-Sun region observed by SUMER, with the
coronal magnetic-field structures as reconstructed from a
simultaneous photospheric magnetogram by means of a force-free-field
extrapolation.} {We show for the first time that coronal funnels
also exist in the quiet Sun. The region studied contains several
small funnels that originate from network lanes, expand with height
and finally merge into a single wide open-field region. However, the
large blue shifts of the Ne~{\sc{viii}} line are not generally
associated with funnels. A comparison between the projections of
coronal loops onto the solar x-y-plane and the Ne~{\sc{viii}}
dopplergram indicates that there are some loops that reveal large
Ne~{\sc{viii}} blue shifts in both legs, and some loops with upflow
in one and downflow in the other leg.} {Our results suggest that
strong plasma outflow, which can be traced by large Ne~{\sc{viii}}
blue shift, is not necessarily associated with the solar wind
originating in coronal funnels but appears to be a signature of mass
supply to coronal loops. Under the assumption that the measured
Doppler shift of the Ne~{\sc{viii}} line represents the real outflow
velocity of the neon ions being markers of the proton flow, we
estimate the mass supply rate to coronal loops to be about
$10^{34}$s$^{-1}$.}
   \authorrunning{Tian et al.}
   \titlerunning{Signature of mass supply to quiet coronal loops}
   \maketitle

\section{Introduction}

Information on the physics of the solar transition region and corona
can be extracted from the solar ultraviolet emission. Since the
plasma is optically thin for most of the emission lines, their
profiles indicate the physical conditions prevailing in the emission
regions. For instance, the line-of-sight velocity of the plasma can
simply be inferred by using the Doppler shift formula
$v_{los}=c(\lambda-\lambda_{0})/\lambda_{0}$, where $\lambda_{0}$
is the wavelength at rest, $\lambda$ is the observed wavelength of
a line, and $c$ is the speed of light in vacuum.

Owing to the high spectral resolution of the SUMER (Solar
Ultraviolet Measurements of Emitted Radiation) instrument onboard
SOHO (Solar and Heliospheric Observatory), the Doppler shifts of
spectral lines can be measured with an accuracy of about 1-2~km/s
\citep{BrekkeEtal1997,HasslerEtal1999,PeterJudge1999,WilhelmEtal2000}.
Hundreds of emission lines in the SUMER spectral window have been
identified by \cite{CurdtEtal2004}. The
Ne~{\sc{viii}}~($\lambda770$) resonance line
($2s^{2}S_{1/2}-2p^{2}P_{3/2}$) is of particular interest for the
study of the upper solar transition region. Since
\cite{DammaschEtal1999} derived a very accurate rest wavelength of
($770.428 \pm 0.003$)$\AA$ for this line, the Ne~{\sc{viii}} Doppler
shift can be precisely determined and thus has been intensively
studied in the recent past.

Generally, the Ne~{\sc{viii}} line was used to study the origin of
the solar wind, and the blue shift of this line is considered to be
a signature of solar-wind outflow. Sizable patches of large blue
shift were found in the Ne~{\sc{viii}} dopplergrams in coronal holes
\citep{Peter1999, StuckiEtal2000, WilhelmEtal2000, AiouazEtal2005};
they tend to be larger in the darker regions of coronal holes
\citep{XiaEtal2003}. \cite{HasslerEtal1999} studied the relationship
between Doppler shift and chromospheric network and found that a
larger blue shift is closely associated with the underlying
chromospheric network. \cite{TuEtal2005a} then found that the
patches of large Ne~{\sc{viii}} blue shift are connected with
coronal funnels, which were reconstructed by the method of
force-free-field extrapolation.

In the quiet Sun, large Ne~{\sc{viii}} blue shifts were also found
in the network lanes and considered to indicate the sources of the
solar wind \citep{HasslerEtal1999}. However, such different
mechanisms as solar wind outflows, spicules, siphon flows through
loops, nano-flares and explosive events all can cause Doppler shifts
of transition region lines \citep{PeterJudge1999}. Strong adjacent
up and down flows have been detected in magnetically active regions
associated with sunspots \citep{Marsch2004}. We must therefore
carefully check whether or not the surrounding magnetic environment
is suitable for a certain mechanism to take effect on the Sun. To
this end, \cite{HeEtal2007} reconstructed the coronal magnetic field
in the quiet Sun by the method of force-free-field extrapolation,
and they found that most of the sites with Ne~{\sc{viii}} blue shift
were not located in regions with an open magnetic field. Thus they
claimed that these sites may not be sources of the solar wind. They
also found that, one dark area on the Fe~($\lambda195$) radiance map
seemed to be connected with open field lines, and conjectured that
it could be a source of the solar wind.

In this paper, we will present new results on the relationship
between the coronal magnetic field and the Ne~{\sc{viii}} blue
shift. Since open and closed field lines can reach different
heights, we will trace the magnetic field lines going up and down,
while starting at every grid point uniformly distributed on a plane
at different heights in our calculation box, in order to illustrate
the two basic kinds of coronal magnetic structures -- funnels and
loops -- and to establish their association with the Ne~{\sc{viii}}
blue shift.

\section{Data analysis and results}

On 22 September 1996, SUMER observed a middle-latitude quiet-Sun
region from 00:40 to 08:15 UTC. There were three main emission
lines, Si~{\sc{ii}} (153.3~nm), C~{\sc{iv}} (154.8~nm), and
Ne~{\sc{viii}} (77.0~nm) in the selected spectral window. In this
paper, we will concentrate on the Doppler shift of Ne~{\sc{viii}}.
This data set has been intensively studied before by
\cite{DammaschEtal1999}, \cite{HasslerEtal1999}, \cite{Peter2000},
\cite{GontikakisEtal2003}, \cite{TuEtal2005b}, and
\cite{HeEtal2007}. More details about the observations can be found
in these papers.

The standard SUMER procedures for correcting and calibrating the
data were applied. We obtained the Ne~{\sc{viii}} dopplergram by
using the same method as described in \cite{TuEtal2005b}. We used
the magnetogram taken by the Michelson Doppler Imager (MDI) on SOHO
on the same day at 01:39 UTC. The pixel size of MDI is about
$2^{\prime\prime}$. The correction of the magnetogram and the
coalignment of the magnetogram with the SUMER images were carried
out by applying the same method described in \cite{TuEtal2005b}, and
therefore are not addressed here.

In our present study we used the force-free-field extrapolation
method as proposed by \cite{Seehafer1978} to extrapolate the
photospheric magnetic field to 80~Mm above the photosphere. Thus,
the size of the calculation box is
$442^{\prime\prime}\times259^{\prime\prime}\times80$~Mm. Those field
lines reaching heigher than 80~Mm are defined as open, otherwise
they are defined as closed. The description of the extrapolation
process, as well as of the suitability of this method for our study,
can be found in \cite{TuEtal2005b}.

\begin{figure}
\resizebox{\hsize}{!}{\includegraphics{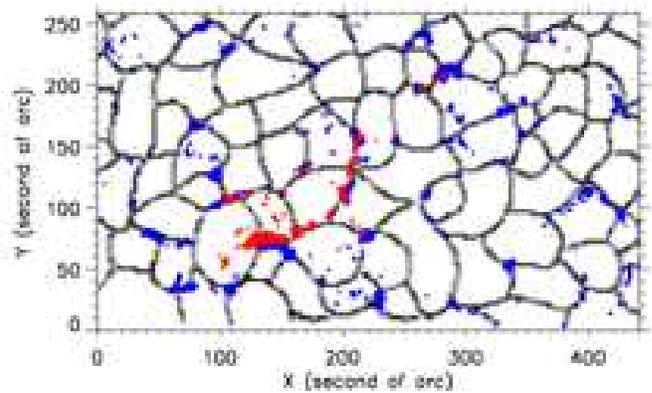}} \caption{Foot
points of magnetic field lines higher than 20~Mm are shown overlaid
on the pattern of the chromospheric network, which was extracted
from a Si~{\sc{ii}} intensity image and is reproduced from the paper
by \cite{HasslerEtal1999}. The red and blue dots denote the foot
points of open and closed field lines, respectively.} \label{fig.1}
\end{figure}

We traced the magnetic field lines going up and down in the corona,
while starting at every grid point uniformly distributed over a
plane located at 20~Mm in our calculation box. Fig.~\ref{fig.1}
shows the distribution of the foot points of these magnetic field
lines. The lanes of the chromospheric network were extracted from a
Si~{\sc{ii}} intensity image and are reproduced from the paper by
\cite{HasslerEtal1999}. It is clear that almost all of the foot
points of the field lines, whether being closed or open, originate
from network lanes and tend to cluster in the network. Since these
dots represent the foot points of field lines reaching higher than
20~Mm, the red and blue clusters of dots can be considered as the
bottom regions of open funnels and large closed loops, respectively.

\begin{figure*}
\sidecaption
\includegraphics[width=13cm]{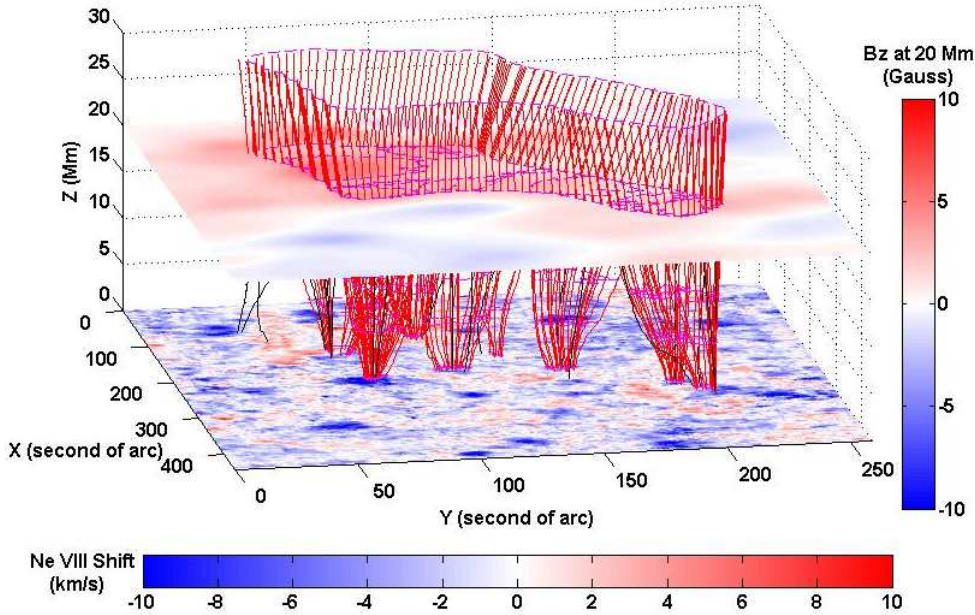}
\caption{~Magnetic funnels in a quiet sun region. The red lines are field lines
originating from the funnel boundary, and the black ones are open field lines
outside small funnels. The Ne~{\sc{viii}} Doppler shift image is placed at
zero megameter, and the color coding is given at the bottom. The positive and
negative values represent red shifts and blue shifts, respectively. The
map of the vertical component of the extrapolated magnetic field at 20~Mm is
placed at the height of 20~Mm, and the color coding is given on the
right-hand bar.}
\label{fig.2}
\end{figure*}

\cite{HeEtal2007} found that in a cross-section plane located at a
height of 25~Mm, the pattern of open field lines intersecting that
plane is consistent with the dark pattern of low radiance in the
image of Fe~{\sc{xii}}~($\lambda 195$). They claimed that this dark
region might be a source of solar wind. Since in coronal holes the
solar wind appears to originate from funnels, it is natural for us
to raise the question whether or not we can also identify funnels on
the basis of the observations of this region. We followed this idea
and reconstructed, for the first time, a coronal funnel structure in
the quiet Sun based on the extrapolation from the photospheric
magnetic field. The results are illustrated in Fig.~\ref{fig.2}.

Each small funnel below 20~Mm originates from a cluster of foot
points of open field lines. These small funnels expand with height
and merge into a single open field region above 20~Mm. The total
area of the cross section at 20~Mm of all the small funnels together
is close to that of the resulting merger funnel. From
Fig.~\ref{fig.1} we can also see some scattered red dots, which are
the foot points of the black field lines in Fig.~\ref{fig.2}. These
lines are also open and go into the funnel merging region, but they
do not originate in any of the small funnels. \cite{AiouazRast2006}
predicted that open flux in the network can be redistributed into
the internetwork region, allowing some of the internetwork field to
remain open in the corona. The ``isolated'' open field lines in our
result seem to originate from the internetwork region, which
confirms this prediction.

In their study, \cite{HeEtal2007} found that most of the sites
showing plasma outflow were not located in regions with an open
magnetic field. From inspection of Fig.~\ref{fig.2} we also find
that most of the funnel bottoms do not coincide with patches of
strong Ne~{\sc{viii}} blue shift. Only one small funnel reveals
considerable blue shift at its bottom (around the coordinates
x=220$^{\prime\prime}$ and y=75$^{\prime\prime}$ in the x-y plane),
while small blue shifts or even red shifts are associated with the
other small funnels.

\begin{figure*}
\sidecaption
\includegraphics[width=13cm]{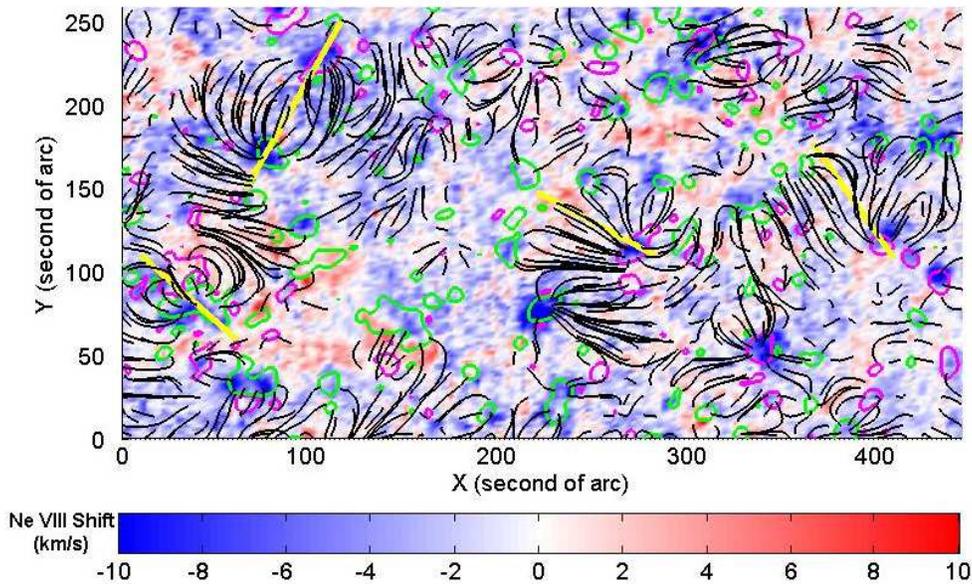}
\caption{~Projections of the extrapolated magnetic loops reaching
higher than 4~Mm onto the x-y-plane, together with the map of the
Ne~{\sc{viii}} Doppler shift. Regions with positive and negative
magnetic flux density are outlined in green and pink color, with the
contour level of 15~G and -15~G, respectively. The four yellow lines
correspond to the four horizontal axes in Fig.~\ref{fig.4}.}
\label{fig.3}
\end{figure*}

As we have now established that large blue shifts in the quiet Sun
mostly do not coincide with funnel bottoms, it seems natural to
consider whether there is some connection between large blue shifts
and coronal loops. Since they are thought to reside mainly in the
lower part of the corona, we traced the magnetic field lines going
up and down, while starting at every grid point uniformly
distributed in a lower plane at 4~Mm in our calculation box. We then
projected the extrapolated magnetic loops reaching higher than 4~Mm
onto the tangent x-y-plane, where the coordinate x refers to
east-west and y south-north direction.

Fig.~\ref{fig.3} shows the results of our correlation study. The
contours of the magnetic field strength coincide well with patches
of large blue shift of Ne~{\sc{viii}}, a similar result to that
previously found by \cite{TuEtal2005b}. However, there clearly are
some cases where both legs of the loop correspond to large blue
shift of Ne~{\sc{viii}}. But there are also some loops with upflow
and downflow in their different legs. It appears as if the magnetic
fields in or around both legs of the loop are strong and comparable,
then the blue shift tends to be strong in both legs. In contrast, if
the magnetic fields are not comparable, one finds blue shift in one
leg and red shift in the other.

\begin{figure*}
\sidecaption
\includegraphics[width=13cm]{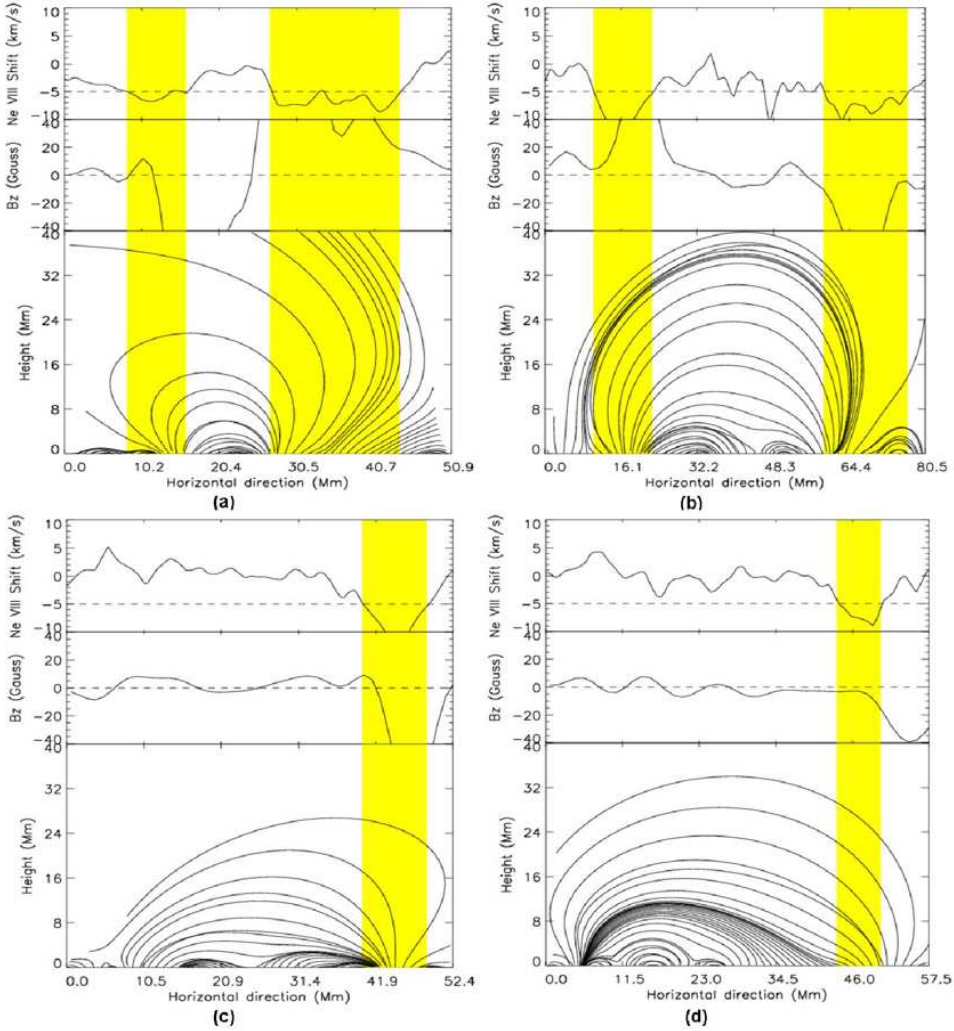}
\caption{~For each subfigure, top panel: the Ne~{\sc{viii}} Doppler
shift; middle panel: the line-of-sight component of the observed
photospheric magnetic field strength along the different cuts
indicated by the yellow lines in Fig.~\ref{fig.3}; bottom panel: the
magnetic field lines projected onto the vertical plane defined by a
cut through the extrapolation box. The four subfigures correspond to
the four cuts (in the sequence from left to right in
Fig.~\ref{fig.3}) and are here indicated by letters (a), (b), (c)
and (d), respectively. The yellow shaded stripes respectively mark
those segments of the plane where the Ne~{\sc{viii}} blue shift is
observed to be strong (unsigned larger than 5~km/s).} \label{fig.4}
\end{figure*}

We plotted the Ne~{\sc{viii}} Doppler shift and the line-of-sight
component of the observed photospheric magnetic field strength along
different cuts, which are shown as yellow lines in Fig.~\ref{fig.3}.
The extrapolated magnetic field lines are projected on the vertical
plane corresponding to each of the four cuts through the
extrapolation box. Fig.~\ref{fig.4} presents the result. For the two
cases shown in the top panel, it is clear that both legs of the big
loop correspond to large Ne~{\sc{viii}} blue shift, and there are
strong magnetic fields (more than 40~gauss) in or around both legs.
For the two cases shown in the bottom panel, only one leg of the big
loop corresponds to a large Ne~{\sc{viii}} blue shift, whereas the
other one corresponds to a red shift. Also, the magnetic field is
asymmetric and different in the two legs; it tends to be much
stronger in or around the leg where the Ne~{\sc{viii}} blue shift
occurs.

\section{Conclusion and discussion}

Stimulated by previous work of \cite{HeEtal2007}, who already
pointed out that most locations with strong blue shifts cannot be
identified with the foot points of open field lines and hence may
not be source regions of the solar wind in the quiet Sun, we further
studied in depth the relationship between strong Ne~{\sc{viii}} blue
shift and the coronal magnetic field structure. We found that there
is a good coincidence of most locations of strong blue shifts with
legs of large coronal loops.

There are also some loops with upflow and downflow in their
different legs, which can be explained by the so-called siphon flow,
where the gas flow is possibly driven by a strong asymmetric heating
in the loop \citep{McClymontCraig1987,Mariska1988,SpadaroEtal1991}.
This type of flow may cause blue shift in one leg and red shift in
the other. The observed asymmetry of the magnetic field strength in
the two loop legs seems to favor such uneven heating. However,
shifts with opposite directions in the two legs may also be caused
by different mechanisms, e.g., with the red shift resulting from any
one of the previously suggested mechanisms
\citep{BrekkeEtal1997,PeterJudge1999}, and with the prominent blue
shift being caused by plasma entering into the loop leg from
outside.

Some of the quiet-Sun-area loops observed here have significant
outflows in both legs, which according to our knowledge, is reported
here for the first time. We suggest that these outflows are caused
by plasma entering into the loop legs from outside by a specific
process, for example, most naturally from the bottom, or perhaps
through side flows, or via reconnection with adjacent field lines of
small neighboring loops, just like in the scenario suggested by
\cite{TuEtal2005a, TuEtal2005c} for the coronal hole funnels. After
having entered the loop legs, the plasma flows up and may accumulate
around the loop top, and then be released to the surrounding corona
when the loop may transiently open through instability or
reconnection.

Our present result revealing that most of the small funnels are not
associated with strong outflow seems puzzling, especially if solar
wind is assumed to flow out along open field lines, as suggested by
\cite{HeEtal2007}. One possible reason is that the outflow in
quiet-Sun coronal funnels is an occasional phenomenon due to
small-scale magnetic activity rather than a continuous one. In fact,
the SUMER raster was done over 7.5 hours, while the MDI magnetogram
reflects the magnetic activity for a short part of this period. Even
though supergranulations have a lifetime of roughly a day ($>7.5$
hours), the network magnetic field is believed to be constantly
reshuffled ($<7.5$ hours), which can produce outflow occasionally
around the relatively unchanged network boundary. However, only
those outflows occurring during an exposure time can be recorded by
SUMER. The fact that one small funnel reveals considerable blue
shift at its bottom (around the coordinates x=220$^{\prime\prime}$
and y=75$^{\prime\prime}$ in the x-y plane) seems to support this
conjecture. However, it must be recognized that the Ne~{\sc{viii}}
blue shift can perhaps not be used as a reliable tracer of solar
wind outflow in the quiet Sun, although this assumption is
considered to be appropriate for coronal holes. Since the magnetic
structures in a coronal hole and the quiet Sun are quite different,
it is likely that in a quiet-Sun region the location where prominent
solar wind outflow starts is higher than the source of the
Ne~{\sc{viii}} emission, and significant outflow could perhaps be
found only in the dopplergram of an ion with a higher formation
temperature than that of the Ne~{\sc{viii}} line. This prediction is
consistent with the results of \cite{TuEtal2005a, TuEtal2005b}, who
obtained a correlation height of 20.6~Mm for Ne~{\sc{viii}} in a
coronal hole and a much lower correlation height of only 3.7~Mm in a
quiet-Sun region. This prediction is hard to be confirmed with the
present data set, because there was no ion with a higher formation
temperature than Ne~{\sc{viii}} in the spectral window covered by
this data. We suggest to employ the emission lines of EIS/HINODE to
check whether prominent outflow can be found in dopplergrams of an
ion formed at a higher temperature than Ne~{\sc{viii}} in the quiet
Sun.

Finally, we will roughly estimate the supply of mass to a coronal
loop under the assumption that the deduced Doppler shift of the
Ne~{\sc{viii}} line represents the real outflow velocity of the neon
ions, which are considered as markers of the proton flow. For a
given loop leg we have, through the mass continuity equation, the
relation:
\begin{equation}
\emph{$f=N_eVA$}\label{equation1},
\end{equation}
where $f$ , $N_e$ ,$V$ and $A$ are the mass flux, electron density,
outflow velocity and area of the loop cross section at the height of
the Ne~{\sc{viii}} emission, respectively.

We took for the electron density in the quiet transition region the
value $N_e=10^{9.64}$cm$^{-3}$ from \cite{GriffithsEtal1999}. Here,
we consider case (b) of Fig.~\ref{fig.4}, for example. First we
selected patches with a blue shift larger than 5~km/s around the two
legs of the loop system on the Ne~{\sc{viii}} dopplergram, and got a
value of 376~arcsec$^2$ and 373~arcsec$^2$ for the two leg areas.
These sizes are slightly larger than the size of a patch of the
chromospheric network lane with a typical value of
$\pi\times10^2$~arcsec$^2$ = 314~arcsec$^2$, and thus can be used as
estimates of the area of the loop cross section at the height of
Ne~{\sc{viii}} emission source. We then averaged the blue shifts in
the two patches, and got average values of 6.92~km/s and 9.26~km/s,
which were used as the average outflow velocities in the two legs of
the loop. Finally, we substituted these parameters into
Eq~(\ref{equation1}), and thus the mass supply rate or mass flux
into the two legs of the coronal loop system can be estimated as
being of the order of $10^{34}$s$^{-1}$.

\begin{acknowledgements}
We thank the anonymous referee for his/her careful reading of the
paper and for the comments and suggestions. We also thank Dr. L.-D.
Xia, C. Zhou and L. Zhao for their initial support of SUMER data
analysis.

H. Tian, C.-Y. Tu, J.-S. He, and G.-Q Zhou are supported by the
National Natural Science Foundation of China under contracts
40574078, 40336053 and 40436015, and by the Beijing Education
Project XK100010404, as well as the foundation Major Project of
National Basic Research contract 2006CB806305. H. Tian is also
supported by China Scholarship Council for his stay in the
Max-Planck-Institut f\"ur Sonnensystemforschung in Germany .

The SUMER project is financially supported by DLR, CNES, NASA, and the ESA
PRODEX programme (Swiss contribution). SUMER and MDI are instruments on
board SOHO, an ESA and NASA mission. We thank the teams of SUMER and
MDI for the spectroscopic and magnetic field data.
\end{acknowledgements}

\end{document}